\documentclass[twocolumn,showpacs,floats,amsmath,amssymb,prb,showkeys]{revtex4}
\usepackage[english]{babel}
\usepackage[dvips]{graphicx}
\usepackage{bm}
\usepackage{dcolumn}
\begin{document}

\title{High-frequency electron paramagnetic resonance investigation of the Fe$^{3+}$ impurity center in polycrystalline PbTiO$_3$ in its ferroelectric phase}

\author{Hrvoje Me\v stri\'c} 
\author{R\"udiger-A. Eichel}
\email[corresponding author, fax: +49-6151-16 43 47, e-mail: ]{eichel@chemie.tu-darmstadt.de}
\author{K.-P. Dinse}
\affiliation{Eduard-Zintl-Institute, Darmstadt University of Technology, D-64287 Darmstadt, Germany}

\author{Andrew Ozarowski} 
\author{Johan van Tol}  
\author{Louis Claude Brunel}
\affiliation{Center for Interdisciplinary Magnetic Resonance, National High Magnetic Field Laboratory, Florida State University, Tallahassee, FL 32310}
\date{\today}

\renewcommand{\thefootnote}{\arabic{footnote}}

\begin{abstract}
The intrinsic iron(III) impurity
center in polycrystalline lead titanate (PbTiO$_3$) was
investigated by means of high-frequency electron paramagnetic
resonance (EPR) spectroscopy in order to determine the
local-environment sensitive fine structure (FS) parameter $D$. At
a spectrometer frequency of 190 GHz, spectral analysis of a powder
sample was unambiguously possible. The observed mean value $D =
+35.28 $ GHz can be rationalized if Fe$^{3+}$ ions substitute for
Ti$^{4+}$ at the B-site of the perovskite ABO$_3$ lattice forming
a directly coordinated Fe$_{\rm Ti}^{\prime}-V_{\rm O}^{\bullet
\bullet}$ defect associate. A consistent fit of the
multi-frequency data necessitated use of a distribution of D
values with a variance of about 1 GHz. This statistical
distribution of values is probably related to more distant defects
and vacancies.\\

\keywords{lead titanate ferroelectrics, iron impurity center, oxygen vacancies, high-field EPR}
\end{abstract}
\pacs{ 61.72.Ji, 61.72.Hh, 61.72.Ww, 76.30.Fc, 77.84.Dy}

\maketitle

\begin{center}
{J. Appl. Phys. Vol. 96, No. 12, 15 December 2004}
\end{center}

\section{Introduction}
Lead titanate (PbTiO$_3$, PT) is widely used as functional ceramic
because of its excellent physical and electromechanical properties
\cite{Jaffe,Xu,Uchino,Lines}. It is  a 'displacive-type'
ferroelectric material that can be used as dielectric in
capacitors and as high refractive index thin film for
electrooptical components. Its piezoelectric and pyroelectric
properties can be utilized in sensors, piezoelectric actuators and
detectors for infrared radiation. As compared to the solid
solution system lead zirconate titanate(Pb[Zr$_x$Ti$_{1-x}$]O$_3$,
PZT), lead titanate exhibits a higher Curie temperature ($T_C =
763$ K) and a lower dielectric constant of about 200, rendering it
more attractive for high-temperature and high-frequency transducer
applications.

In order to improve material properties, several transition metals
or rare-earth elements may be added on a percentage level, for
which reason considerable interest exists to characterize the role
of such extrinsic functional centers. Generally, standard bulk
characterization techniques fail due to the inherent low
concentration of these centers and electron paramagnetic resonance
(EPR) becomes the {\it method-of-choice} because of its high
sensitivity and selectivity \cite{PT_EPR}. In particular, if
doping with iron is considered, the local symmetry can be explored
by monitoring the resulting fine structure (FS) interaction, which
will be modified by the presence of oxygen vacancies ($V_{\rm
O}^{\bullet \bullet}$)\cite{Kroeger}. If charge compensation occurs in the
nearest-neighbor O$^{2-}$ ion shell, a large distortion of the
octahedral symmetry results and because of the short distance a
large change of the intrinsic Fe$^{3+}$ FS tensor will be
induced \cite{Intrinsic}.

Unless crystalline samples are available, these tensor elements in
general are not directly accessible at X-band (9.4 GHz)
frequencies, since the resulting zero-field splitting (ZFS) is
much larger than the microwave (mw) quantum energy. FS values
therefore have to be deduced from the analysis of second-order
line shifts in a fully resolved single crystal spectrum. If only
powder samples are available, high-frequency EPR has to be invoked
in order to approach high-field conditions, under which an
accurate determination of the principal value $D$ of the FS
interaction (including sign) via first-order effects is
possible.

Beyond its technical importance, we can use lead titanate as a
model system related to the considerably more complex
acceptor-doped PZT system \cite{PZT_EPR}, because its structure
being comparatively simple and well defined. Furthermore, a
prerequisite for an analysis of the EPR spectra of solid-solution
PZT systems with varying Pb/Zr composition is a complete
characterization of the spectra of pure compounds. As stated
above, in general it is quite difficult to extract large ZFS
tensor elements from X-band EPR powder spectra. In order to
establish high-frequency EPR as reliable tool for the
investigation of technologically relevant polycrystalline
compounds, we have chosen the ubiquitous  Fe$^{3+}$ impurity
center in lead titanate powder as first example, because FS
parameters have been obtained by previous single crystal studies
which can be used for comparison.

The first EPR spectrum, attributed to Fe$^{3+}$ in PbTiO$_3$ led
to the discovery of an unusual high FS splitting \cite{Gai64}.
This observation was related to the strong tetragonal
ferroelectric distortion of the crystalline field. At Q-band (35
GHz), two different iron centers were observed, one of them being
assigned to a partially charge-compensated Fe$_{\rm
Ti}^{\prime}-V_{\rm O}^{\bullet \bullet}$ defect associate. It was
also observed that the FS interaction increases by 20 \% when
cooling from room temperature to 77 K, consistent with a change in
lattice deformation, the $c/a$ ratio changing by 19 \% in this
temperature range. The spectra allowed also for the determination
of additional terms attributed to higher-rank tensor elements. The
temperature dependence of line intensities at liquid helium
temperatures indicated that the sign of $D$ is positive
\cite{Pon69}. At 70 GHz, additional transitions could be observed
and allowed to refine the ZFS parameters \cite{Lew76}. In contrast
to these findings, in a recent X-band EPR study  only a single
Fe$^{3+}$ center was observed, for which charge compensation was
assumed to occur at distant spheres \cite{Lag96}. This discrepancy
was attributed to different synthesis techniques involved in
manufacturing the crystals. At low temperatures hyperfine
structure, assigned to interaction with nearby $^{207}$Pb nuclei
was observed.

The reported ZFS parameters are spread over a broad range. Since
many attempts for gaining structural information about the dopant
site, such as the assignment to either Fe$_{\rm
Ti}^{\prime}-V_{\rm O}^{\bullet \bullet}$ defect associates or
not-coordinated 'free' Fe$_{\rm Ti}^{\prime}$ centers with or
without off-center shifts of the iron ion, are based upon size and
sign of the ZFS parameters, their precise determination is of
considerable importance. By taking the reported interval of ZFS
parameters as input parameters for modelling the structure of the
iron center, a contradictory variety of proposed structures
results, ranging from 'free' Fe$_{\rm Ti}^{\prime}$ centers to
Fe$_{\rm Ti}^{\prime}-V_{\rm O}^{\bullet \bullet}$ defect
associates with off-center shifts towards or away from the oxygen
vacancy. In this work, we hence aim for an unambiguous
determination of the ZFS parameter by application of
high-frequency EPR up to 190 GHz, thus creating the basis for an
accurate modelling of the structure. In the magnetic field range
of 2 to 8 T used, the electron Zeeman energy will be the dominant
term in the spin Hamiltonian. Resulting quantization approximately
along the field direction significantly simplifies the EPR
spectrum. Accordingly it is possible to accurately determine  ZFS
parameters of Fe$^{3+}$ in polycrystalline PbTiO$_3$ from observed
van Hove singularities, which are related to canonical
orientations of the compound. High frequency EPR thus is an
indispensable tool for the investigation of technologically
relevant PZT, which are commonly available only as disordered
compounds.

\section{Experimental}
9.5 GHz continuous wave (c.w.) EPR measurements were performed
using a ESP 300E spectrometer (Bruker), equipped with a
rectangular TE$_{112}$ resonator. The magnetic field was read out
with a NMR gaussmeter (ER 035M, Bruker) and as a standard field
marker polycrystalline DPPH with $g = 2.0036$ was used for the
exact determination of the resonance magnetic field values.
High-frequency EPR measurements were performed at the National
High Magnetic Field Laboratory (NHMFL) Tallahassee \cite{Has00}.
The set-up used operates in transmission mode and employs
oversized cylindrical wave guides. No resonator was used.
Microwave detection was performed with a low-noise, fast response
InSb hot-electron bolometer (QMC Ltd.), operated at liquid-helium
temperature. Field modulation in the range of 10 to 50 kHz was
used to obtain 'first-derivative' type EPR spectra.

\section{Theoretical description}
In principle, a discussion of the Fe$^{3+}$ center in its high-spin form $(S = \frac{5}{2})$
should be based upon the most general form of the spin
Hamiltonian, including fourth-rank tensor components \cite{Abr70}.
However, literature values of spin-Hamiltonian parameters for
Fe$^{3+}$ ions in lead titanate single crystals, as well as our
high-field EPR results justify considerable simplifications.
First, since no significant line broadening was observed in high
magnetic fields, the anisotropy of the $g$-matrix apparently
imposes a much smaller angular-dependent EPR resonance shifts than
the inherent peak-to-peak linewidth $\Delta B_{\rm pp}$ of the
resonance lines and might therefore be neglected. This is in
agreement with expectation because of a vanishing orbital momentum
of Fe$^{3+}$ ions. As a consequence, the electron Zeeman
interaction will be represented by an isotropic $g$-value. Second,
due to the relatively low natural abundance of the only magnetic
active isotope $^{57}$Fe, iron hyperfine interaction will be
omitted for spectrum reconstruction. Third, the nuclear Zeeman interaction is ignored as well because it does not contribute to EPR spectral features in first order. Finally, no influence
of fourth-rank tensor elements could be experimentally resolved.
Hence, the corresponding terms were neglected.

\begin{figure}
   \includegraphics[width=0.9\linewidth]{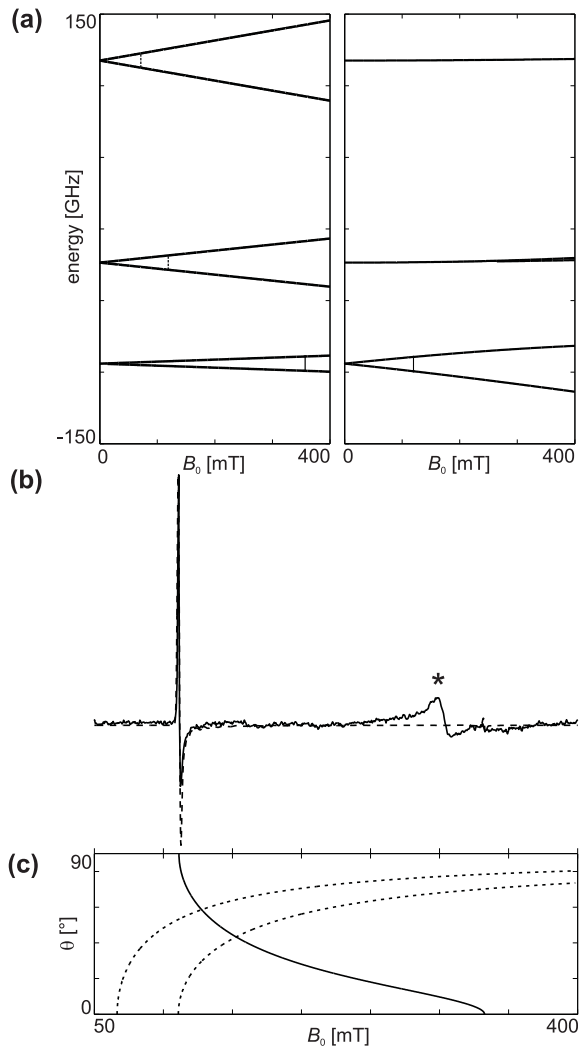}
  \caption{\label{X_band}(a) Schematic representation of energy levels as function of magnetic field for the canonical parallel (left) and perpendicular (right) orientations relative to the ZFS-tensor principal
  axes for the Fe$^{3+}$ impurity center in PbTiO$_3$. The FS interaction is taken to be
  positive, $D > 0$. EPR transitions for X-band frequencies are marked by vertical lines
  involving 'allowed' (solid) and 'forbidden' transitions (dashed). (b) X-band EPR spectrum
  of the Fe$^{3+}$ impurity center in PbTiO$_3$ at $\nu_{\rm mw} = 9.320$ GHz and
  at 10 K. The line at $g = 2.2$ marked by an asterisk is tentatively assigned to Pb$^{3+}$ centers.
  (c) Orientation dependence of the resonance lines for 'allowed' and 'forbidden' transitions.}
\end{figure}

The resulting simplified effective spin Hamiltonian is given by
\begin{equation}
 \label{Hamiltonian}
 {\cal H} = {\bf S \cdot D \cdot S} + g_{\rm iso} \beta_e {\bf B_0 \cdot S}
\end{equation}
in which $g_{\rm iso}$, $g_n$ are the electron and nuclear $g$
factors and $\beta_e$, $\beta_n$ are the Bohr and nuclear
magnetons. The second and third terms are the electronic and
nuclear Zeeman interactions, respectively, and $\bf B_0$ is
denoting the external field. The first term describes the
second-rank fine structure interaction. This term is often
referred to as the zero-field splitting term because it lifts the
degeneracy of energy states independent of magnetic field. The FS
tensor $\bf D$ is symmetric and traceless. Thus, there is always a
coordinate system in which the tensor is diagonal with the
elements $D_{\parallel}$ and $D_{\perp}$. By convention,
$D_{\parallel}$ is taken to be the principal value with the
largest absolute magnitude. Hence, for axial symmetry, the
fine-structure splitting may be described in terms of a single
crystal-field parameter that is commonly defined as $D =
\frac{3}{2} D_{\parallel} = 3 D_{\perp}$. The free Fe$^{3+}$ ion
has a $3d^5$ $^6S_{5/2}$ ground state. Under the action of a
tetragonal crystalline field, this ground state splits into three
twofold degenerate levels (cf. figure \ref{X_band} (a)). 

In case of the commonly employed X-band frequencies, the
zero-field splitting is much larger than the electron Zeeman
energy. All doublets are well separated and only transitions
within each doublet can be observed. Accordingly, the spectrum
then is described by defining a {\it separate} effective spin
Hamiltonian with $S' = \frac{1}{2}$ for each doublet. Because of
orientation-dependent state mixing, the effective $g'$-matrices
can exhibit a very large anisotropy even retaining the assumption
that the pure electron Zeeman interaction is isotropic
\cite{g_eff}. This situation is schematically represented in
figure \ref{X_band}(a, c). At zero field, the spins are quantized
along the $z$-axis forming three groups of doubly degenerate
states. After application of a static field along the $z$-axis,
the quantization along the crystal $z$-axis is retained, the
degeneracy being lifted proportional to $B_0$. However, due to the
zero-field splitting being larger than the mw quantum of energy at
X-band frequencies, only transitions between levels degenerate at
zero field may be induced, providing such transitions are not
'forbidden' by the selection rule $\Delta m_S = \pm 1$. If the
magnetic field is perpendicular to the $z$-axis, there is no
first-order splitting of the $| m_S = \pm \frac{5}{2} \rangle$ and
$| \pm \frac{3}{2} \rangle$ states, whereas the $| \pm \frac{1}{2}
\rangle$ states are split with $g' \approx 6$. Due to the
selection rule all transitions, except the $| m_S = + \frac{1}{2}
\rangle \leftrightarrow | - \frac{1}{2} \rangle$ transition are
'forbidden' and will not be observed in the X-band EPR spectrum.

\section{Results and discussion}
A typical X-band EPR spectrum is presented in figure
\ref{X_band}(b). It mainly consists of one prominent feature at
low field. In terms of effective $g$-values, the spectrum is
described  with $g'_{\parallel} = 2.004$ and $g'_{\perp} = 5.956$.
Because  all orientations are statistically realized in a powder,
the resonances extend from $g'_{\parallel}$ to $g'_{\perp}$, as
illustrated in figure \ref{X_band} (c), and because the
probability of $\bf B_0$ being perpendicular to the $z$-axis is
largest, the resonance peak at $g'_{\perp}$ is the dominant
feature. The additional feature at $g = 2.2$ can be assigned to
Pb$^{3+}$ centers \cite{Pb3+}.

\begin{figure}
  \includegraphics[width=0.9\linewidth]{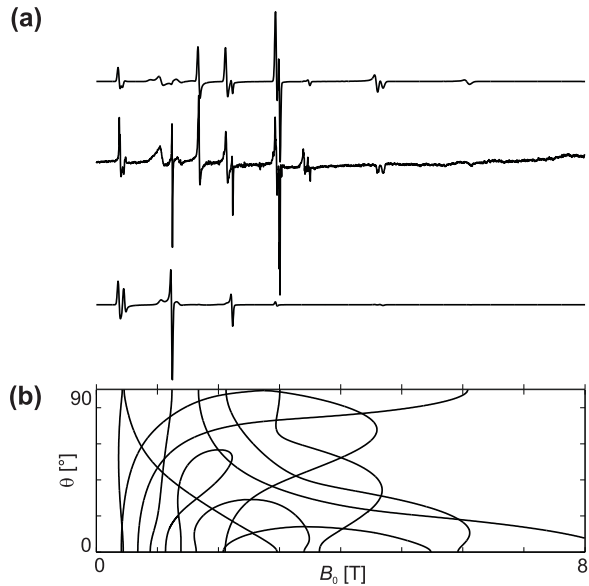}
  \caption{\label{W_band}(a) W-band EPR spectrum of the Fe$^{3+}$ impurity center in PbTiO$_3$ at
  $\nu_{\rm mw} = 95.284$ GHz (T = 10 K). Numerical spectrum simulations
  involve $\bf B_1$ in perpendicular (top) and parallel (bottom) mode. (b) Orientation
  dependence of EPR transitions.}
\end{figure}

The W-band EPR spectrum (95 GHz), depicted in figure
\ref{W_band}(a), is representative for an intermediate field
regime. For this situation the crystal-field terms are comparable
to the electron Zeeman interaction. Hence, no simple prediction
pattern of the EPR spectrum results. The observed sharp lines,
however, can be identified as arising from to resonances with
minor orientation dependence, as shown in figure \ref{W_band}(b).
To obtain good agreement between simulation \cite{simul} and experiment, it had
to be assumed that the sample was exposed to oscillating $\bf B_1$
field components being both parallel and perpendicular to $\bf
B_0$. This phenomenon is caused by multiple reflections and
general imperfections (mode impurity) in the mw propagation
system, in contrast to cavity-based systems with a well-defined
$\bf B_1$ polarization \cite{Krz04}. For spectrum simulations this
effect necessitates consideration of parallel and perpendicular
EPR modes. In the perpendicular mode, the excitation and detection
mw fields are along the laboratory $x$-axis, whereas in the
parallel mode they are along the $z$-axis, parallel to the
external static field. When the magnetic field is parallel to a
principal axis, the EPR $\Delta m_S = \pm 2$ transition
probability drops to zero with the modulation field $\bf B_1$
perpendicular to $\bf B_0$, while it is largest with $\bf B_1$
parallel to $\bf B_0$, enabling the observation of 'forbidden'
transitions.

\begin{figure}
  \includegraphics[width=0.9\linewidth]{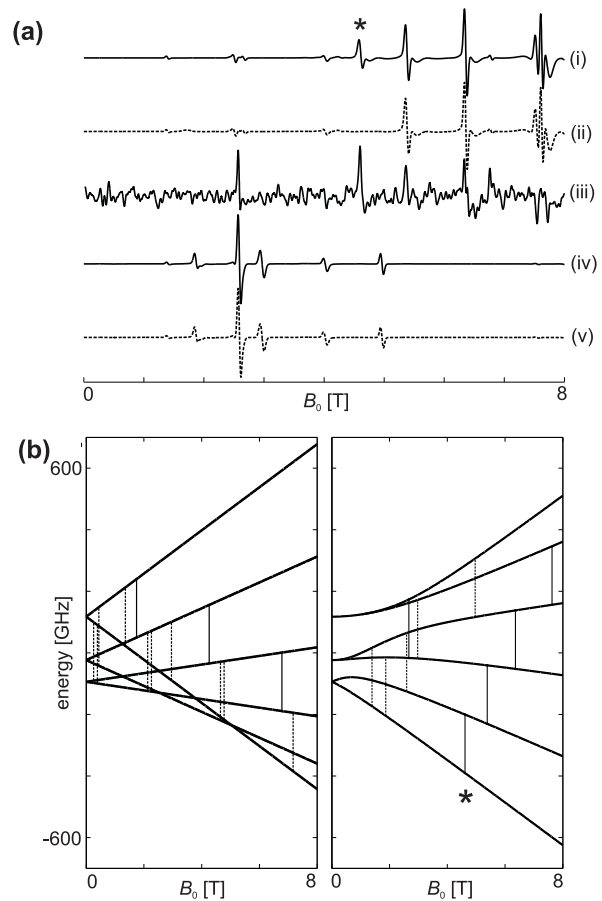}  
  \caption{(a) G-band EPR spectrum of the Fe$^{3+}$ impurity center in PbTiO$_3$ at
  $\nu_{\rm mw} = 189.962$ GHz and T = 10 K (iii). Numerical spectrum simulations
  involve both $\bf B_1$ in perpendicular (i,ii) and parallel (iv,v) mode, as well as
   a positive (solid) and negative sign (dashed) of the axial ZFS parameter. The transition
   being particularly sensitive to temperature-dependent level population is marked by an
   asterisk. (b) Energy levels as function of magnetic field for the canonical
   parallel (left) and perpendicular (right) orientations relative to the ZFS-tensor principal
   axes using $D > 0$. EPR transitions for  G-band frequencies are marked by vertical
   lines involving 'allowed' (solid) and 'forbidden' transitions (dashed).}
  \label{G_band}
\end{figure}

At G-band (190 GHz), EPR is performed almost in the high-field
regime. The corresponding energy-level diagrams for the canonical orientation
perpendicular relative to the ZFS-tensor principal axes is depicted
in figure \ref{G_band}(a). The field-energy dependencies are quite
complicated because of extensive level mixing, resulting in
strongly varying resonance fields for most of the transitions.
The EPR spectrum is presented in figure \ref{G_band}(b). At low temperatures, the sign of the $D$ can in principle be
determined. This is due to the fact that the transition
intensities are a function of the Boltzmann populations of the
levels involved. In our case FS splitting and electron Zeeman
energies are much larger than $k_B T$ at 10 K at which temperature spectra were
recorded for the determination of the sign of $D$. Under these conditions, the temperature  dependence of particular
resonances allows to determine the sign of the  splitting
parameter $D$. In figure \ref{G_band} (a), results for numerical
spectrum simulations assuming positive (solid) and negative sign
(dashed) of $D$ are shown. Because the resonance at 4.61 T (marked
by an asterisk) occurs in the calculated spectrum only for $D >
0$, the ZFS interaction can be taken as positive. The disappearance
of the signal for $D < 0$ can be understood by taking into account
the calculated energy level diagrams in figure \ref{G_band}(a),
where the particular line corresponds to the transition connecting
states of lowest energy. If the sign of $D$ would be reversed,
this level would be highest in energy and hence this transition
has to be thermally activated and should be detectable at elevated
temperatures only.

\begin{table*}
 \label{results}
 \caption{Table I: {$g$-values and FS parameters as obtained from high-field EPR, compared to values from literature with $\nu_{\rm mw}$. The sign of the principal ZFS parameter is determined by an analysis of line intensities at low temperatures. (For a definition of the fouth-order parameters $a$, $F$ see for example: A. Abragam, B. Bleaney: {\it Electron Paramagnetic Resonance of Transition Ions}, Clarendon Press, Oxford (1970))}}
 \begin{center}
 \begin{tabular}{|l|c|c|c|c|r|r|r|} \hline
  center & $g_{\rm iso}$ & $D$ [GHz] & $a$ [MHz] & $F$ [MHz] & $T$ [K] & $\nu_{\rm mw}$ [GHz] & \\ \hline \hline
  Fe$^{\prime}-V^{\bullet \bullet}_{\rm O}$ & 2.002 & $35.28 \pm 0.01$ & & & 2-10 & 9.6 - 190 & this work \\ \hline
  Fe$^{\prime}$ & 2.009(5) & $27.13 \pm 0.01$ & $1678 \pm 10$ & $-2833 \pm 10$ & 290 & 9.4 & \cite{Lag96} \\
  & & $32 \pm 1$ & & & 77 & 9.4 & \\ \hline
  Fe$^{\prime}-V^{\bullet \bullet}_{\rm O}$ & & $35.59 \pm 0.06$ & $600 \pm 210$ & $-510 \pm 300$ & 85 & 70 & \cite{Lew76} \\
  Fe$^{\prime}$ & & $34.48 \pm 0.09$ & $660 \pm 300$ & $-600 \pm 300$ & & & \\ \hline
  Fe$^{\prime}-V^{\bullet \bullet}_{\rm O}$ & & $26.98 \pm 1.5$ & & & 290 & 37.5 & \cite{Pon69} \\
  Fe$^{\prime}$ & 2.0 & $15.89 \pm 0.6$ & $809 \pm 120$ & & & & \\
  Fe$^{\prime}-V^{\bullet \bullet}_{\rm O}$ & & $32.98 \pm 1.5$ & & & 77 & & \\
  Fe$^{\prime}$ & 2.0 & $20.08 \pm 0.6$ & & & & & \\ \hline
  Fe$^{\prime}$ & 2.009(5) & $\ge$ 30 & & & & 9.4 & \cite{Gai64} \\ \hline
 \end{tabular}
\end{center}
\end{table*}

The refinement of the spin-Hamiltonian parameters through numerical
spectrum simulation simultaneously for all mw frequencies led to
the results summarized in table I. Using these
best-fit values, numerically simulated spectra are superimposed to
the experimental data and show excellent agreement. Concerning the
relatively large value of $D$, the local environment for the
Fe$^{3+}$ ion can be modelled with the help of the semi-empirical
{\it Newman superposition model} \cite{NSM1}. The geometry of the
next-nearest oxygen positions was calculated using the lattice
parameters $c = 0.422$ nm and $a = 0.386$ nm from XAFS
measurements at 12 K \cite{NSM2}. The parameters for the
iron-oxygen ion pair were taken as $\overline{b}_2 = -12.36$ GHz
at a reference distance of $R_0 = 0.2101$ nm, and the power-law
exponent as $t_2 = 8$ \cite{NSM3}, respectively. Only by assuming
Fe$^{3+}$ being located at the B-site correlated with a
next-nearest oxygen vacancy, the measured value for the ZFS could
be reproduced within the centered model \cite{NSM4}. An off-center
displacement of the iron in an intact oxygen octahedron leads to a
considerably smaller value ($D \le 25$ GHz). Based upon the
experimentally determined FS value, the conclusion is thus drawn
that iron builds a defect associate with a directly coordinated
oxygen vacancy (Fe$'_{\rm Ti}-V_{\rm O}^{\bullet \bullet}$).
These results are in agreement with predictions obtained by
ab-initio calculations, which are in progress.

Moreover, it can be concluded from the axial symmetry of the ZFS tensor that the orientation of the Fe$'_{\rm Ti}-V_{\rm O}^{\bullet \bullet}$ defect dipole is along the crystallographic $c$-axis. Any other coordination of the oxygen vacancy along either $a$ or $b$ would reduce the symmetry of the ZFS tensor to orthorhombic, which was not observed within spectral resolution.

Starting with $D$ as a single fit parameter, a satisfactory
reproduction of the experimental data taken at all different mw
frequencies was possible only when allowing for statistical
Gaussian distributions $\delta D$. Considering the
inhomogeneous charge-compensation mechanisms, provided by either
lead vacancies ($V'_{\rm Pb}$) or valency-altered Pb$^{3+}$
centers, and considering that these defects can be in different
shells of the Fe$^\prime_{\rm Ti}-V_{\rm O}^{\bullet \bullet}$
defect associate, a statistical distribution (strain) of the
spin-Hamiltonian parameters defined by corresponding variances is
expected. In our case, a variance $\delta D = 1$ GHz of the ZFS
parameter $D$, taken to be Gaussian and independent over the mw
frequency range from 9.6 to 94 GHz, had to be assumed.

The high-field approach thus allows for determining absolute value
and sign of the axial fine structure parameter for polycrystalline
compounds without ambiguity and even superior accuracy as compared
to single crystal studies at X-band frequencies. This is due to
the fact that FS values at high fields can accurately be
determined via first-order effects rather than by second-order
shifts at low fields. A second consequence is that a distribution
$\delta D$ of the FS parameter is observable as first-order
variation of the line positions that result in line broadening in
the high-field spectra. At X-band, this effect is of second-order
only and thus almost vanishes. Hence no information about
$D$-strain can be gathered. In contrast, fourth order FS
parameters were not accessible by numerical spectrum simulations
up to 190 GHz, which in turn are accessible by
orientation-dependent single-crystal studies at X-band \cite{Pon69,Lew76,Lag96}.

\section{Acknowledgments}
We are grateful to Dr. Theo Woike for providing the sample. This research has been financially supported by the DFG priority program 1051 {\it 'High-Field EPR in Biology, Chemistry and Physics'} and center of excellence 595 {\it 'Electrical Fatigue in Functional Materials'}. The NHMFL is funded by the NSF through Grant DMR9016241. K.P.D. is grateful for a visiting professor fellowship at the NHMFL.


\begin{thebibliography}{99}
\bibitem{Kroeger} The Kr\"oger and Vink notation is used to describe the charge state of the defect with respect to the neutral lattice.
\bibitem{Intrinsic} 'Intrinsic' in this context refers to the fact that the B-site symmetry in the ferroelectric phase already allows the existence of an axial FS tensor even without the presence of oxygen vacancies.
\end{thebibliography}
\end{document}